\documentclass[twocolumn,showpacs,amsmath,amssymb, superscriptaddress]{revtex4}

\usepackage{amsmath}
\usepackage{amssymb}
\usepackage{amsfonts}

\usepackage{graphicx}
\usepackage{float}
\usepackage{color}

\usepackage{amsthm}

\DeclareMathOperator{\Tr}{Tr}

\DeclareMathOperator{\su}{\mathfrak s\mathfrak u}

\newcommand{\paren}[1]{{\left( #1 \right)}}

\renewcommand{\Re}{{\rm Re}}

\newcommand{\pder}[2]{{\partial{#1}\over\partial{#2}}}

\makeatletter
\def\loweq@align#1#2{\lower.6ex\vbox{\baselineskip\z@skip\lineskip\z@
    \ialign{$\m@th#1\hfil##\hfil$\crcr#2\crcr=\crcr}}}
\def\lowsim@align#1#2{\lower.6ex\vbox{\baselineskip\z@skip\lineskip\z@
    \ialign{$\m@th#1\hfil##\hfil$\crcr#2\crcr\sim\crcr}}}
\def\geqq{\mathrel{\mathpalette\loweq@align >}}
\def\leqq{\mathrel{\mathpalette\loweq@align <}}
\def\grsim{\mathrel{\mathpalette\lowsim@align >}}
\def\lesssim{\mathrel{\mathpalette\lowsim@align <}}
\def\gsim{\mathrel{\mathpalette\lowsim@align >}}
\def\lsim{\mathrel{\mathpalette\lowsim@align <}}
\makeatother
\newcommand{\grless} 
{ {\, \raise-.24em\hbox{$<$} \hspace{-0.8em} \raise.31em\hbox{$>$}\, } }
\newcommand{\lessgr} 
{ {\, \raise-.24em\hbox{$>$} \hspace{-0.8em} \raise.31em\hbox{$<$}\, } }
\newfont{\bg}{cmr10 scaled\magstep4}                    
\newcommand{\bigzerou}{\smash{\lower1.7ex\hbox{\bg 0}}}

\newcommand{\nn}{\nonumber \\ }

\newcommand{\crl}[1]{[-\infty,\infty]}

\newcommand{\ket}[1]{|{#1}\rangle}

\newcommand{\bra}[1]{\langle{#1}|}

\newcommand{\Ref}[1]{(\ref{#1})}

\newcommand{\wt}{\widetilde}

\newcommand{\da}[1]{#1^\dag}

\newcommand{\vv}[1]{{\boldsymbol{#1}}}

\newcommand{\dd}{{\rm d}}

\begin{document}

\title{Quantum Brachistochrone for Mixed States}
  \author{Alberto Carlini}
 \email{carlini@ics.mq.edu.au}
 \affiliation{Centre for Quantum Computer Technology, Department of Physics, Macquarie University, Sydney, Australia}
\author{Akio Hosoya}
 \email{ahosoya@th.phys.titech.ac.jp}
 \affiliation{Department of Physics, Tokyo Institute of
 Technology, Tokyo, Japan}
\author{Tatsuhiko Koike}
 \email{koike@phys.keio.ac.jp}
 \affiliation{Department of Physics, Keio University, Yokohama, Japan}
\author{Yosuke Okudaira}
 \email{okudaira@th.phys.titech.ac.jp}
 \affiliation{Department of Physics, Tokyo Institute of
 Technology, Tokyo, Japan}


\begin{abstract}

We present a general formalism based on the variational principle for finding the time-optimal quantum evolution of mixed states governed by a master equation, when the Hamiltonian and the Lindblad operators  are subject to certain constraints.  
The problem reduces to solving first a fundamental equation (the {\it quantum brachistochrone}) for the Hamiltonian, which can be written down once the constraints are specified, and then solving the constraints and the master equation for the Lindblad and the density operators. 
As an application of our formalism, we study a simple one-qubit model where the optimal Lindblad operators control decoherence and can be simulated by a tunable coupling with an ancillary qubit. 
It is found that the evolution through mixed states can be more efficient than the unitary evolution between given pure states.
We also discuss the mixed state evolution as a finite time unitary evolution of the system plus an environment followed by a single measurement.
For the simplest choice of the constraints, the optimal duration time for the evolution is an exponentially decreasing function of the environment's degrees of freedom.

\end{abstract}  


\maketitle

\section{Introduction}

Recently, many works related to time optimal quantum computation have appeared in the literature \cite{khaneja,khaneja1,vidal,zhang,tanimura,schulte,boscain,nielsen1,nielsen3,khanejanew,dowling}. 
The minimization of physical time to achieve a given unitary transformation should provide a more physical description of the complexity of quantum algorithms.
In a series of previous works \cite{paper1, paper2} we established a general theory based on the variational principle to find the time optimal evolution between given initial and final pure states 
\cite{paper1} (paper I), and to find the time optimal way of generating a target unitary operation for arbitrary initial states \cite{paper2} (paper II).
In paper I we studied closed pure quantum systems driven by the Schr\"odinger equation and where the Hamiltonian is controllable within a certain available set. 
Paper II is an extension of paper I and is more relevant to subroutines in quantum computation, where the input may be unknown.
The main result of our works is that, once the constraints for the Hamiltonian are given, one can systematically derive a fundamental equation, the {\it quantum brachistochrone}, which can be always
solved, at least numerically, for the time optimal Hamiltonian (without any further restricting assumptions, e.g.  the adiabaticity of the quantum evolution).
Here we extend our previous works 
and formulate a variational principle for the time optimal quantum control
of open systems where the dynamics is driven by a master equation in the Lindblad 
\cite{lindblad,gorinikossakowskisudarshan} form.

Historically, quantum control theory of pure states has been studied by many
people (for a review of the subject, see, e.g., \cite{shapirobrumer}). 
Around twenty years ago, Peirce, Dahleh and Rabitz \cite{peircedahlehrabitz} considered a variational method to manufacture a wave packet as close as possible to a given target.
Other authors (see, e.g., \cite{tannorrice} and other references in \cite{shapirobrumer}) further investigated the variational methods by optimizing the fidelity between the final state of the steered system and a given target state.
The application to the optimal realization of unitary gates in closed systems was also studied (see, e.g., 
\cite{palao}).
For the mixed state case, the master equation in the Lindblad form has been used in \cite{lloydviola} 
(for other recent references see, e.g.,  \cite{grigorenko,schneider,schulte2}). 
However, while the main concern of these papers was the optimization of the fidelity or the purity of the quantum operations, here we focus the attention on the time optimality. 

The paper is organized as follows.
In Sec. ~\ref{sec-master} we introduce the Markovian approximation and the master equation for
an open quantum system and we discuss the related gauge degrees of freedom.
In Sec.~\ref{general formalism} we set up the general variational formalism for the time optimal evolution of such quantum systems. 
The action consists of a time cost function to be minimized and of Lagrange multiplier terms which ensure the evolution under the master equation and certain constraints for the available Hamiltonian and Lindblad operators.
We then derive the fundamental equations of motion.
In Sec.~\ref{onequbit} we explicitly demonstrate our methods via the example of 
a one-qubit system by deriving the time optimal Hamiltonian, Lindblad and density
operators, which may represent either a measurement or a decoherence process. 
In Sec.  ~\ref{twoqubits} we simulate the optimal operations derived in Sec. ~\ref{onequbit} 
by the partial trace of a two-qubit self-interacting system with a controllable Hamiltonian and ancillary qubit. This corresponds to a repetition of short-time measurements.  
Sec. ~\ref{closedsystem}, instead,  describes the mixed state evolution as a finite time unitary evolution of the system plus an environment followed by a single measurement. 
Finally, Sec.~\ref{summary} is devoted to the summary and discussion of our results.

\section{Master equation}
\label{sec-master}
\newcommand{\Chi}{{\boldsymbol\chi}}
\newcommand{\LL}{{\cal L}}
In this paper, 
we address the problem of time optimal quantum control 
of open systems where the dynamics is described by a master equation in the
Lindblad  form 
\cite{lindblad,gorinikossakowskisudarshan} 
\begin{equation}
 \dot\rho :=\mathcal{L}(\rho)
 =-i [H,\rho] + \sum_a\left (L_a\rho \da L_a-\frac{1}{2}\{\da L_aL_a,\rho\}\right )
\label{master}
\end{equation}
for the density operator $\rho(t)$, where $H(t)$ is the Hamiltonian,
$L_a(t)$ ($a=1,..., N^2-1$)
are the Lindblad operators, $N$ is the dimension of the Hilbert space of the system and we use the notation $\dot A:= dA/dt$ for time derivatives, $[A, B]:=AB-BA$ for the commutator and $\{A, B\}:=AB+BA$ for the anticommutator.
The Hamiltonian represents the unitary evolution part while the Lindblad operators express generalized measurements or decoherence processes due to the coupling of the system with an environment.
The master equation is Markovian, i.e., 
the environment has no memory of the main physical system. 
It can be physically
realized if the interaction between the main system and 
the environment is weak and its
typical time scale is small compared with that of the
physical system ~\cite{petruccione,vanhove}. 
A simple example discussed in Sec. V illustrates how the repetition of a short time unitary evolution and the partial trace  (e.g., measurement) over the environment states \Ref{2model} reproduces the master equation \Ref{master}.

The evolution of $\rho(t)$ is invariant under the following gauge
transformations
\begin{align}
  &H\mapsto H+\alpha(t)
  +\frac1{2i}\sum_a[\beta_a^*(t)L_a-\beta_a(t)\da L_a], 
 \nonumber \\
  &L_a\mapsto L_a+\beta_a(t), 
    \label{eq-gauge-1}
\end{align}
and
\begin{align}
  \label{eq-gauge-2}
  &H\mapsto H, 
  \quad
  L_a\mapsto \sum_b U_{ab}(t)L_b, 
\end{align}
where $\alpha(t)$ is a real number, 
$\beta_a(t)$ are complex numbers, 
and the $U_{ab}(t)$ form a unitary matrix with respect to its indices (i.e., 
$\sum_cU_{ac}U^{\ast}_{bc}=\sum_cU_{ca}^{\ast}U_{cb}=\delta_{ab}$).
The parameter $\alpha(t)$ in \Ref{eq-gauge-1} corresponds to the $U(1)$ gauge degree of freedom 
which was discussed in our first papers \cite{paper1, paper2} for the case of pure states.
The gauge degrees of freedom $\beta_a(t)$, instead, correspond to the fact that the operator $H(t)$ in \Ref{master} is not just the free Hamiltonian of the reduced system, but may contain coupling terms with the environment (see, e.g., equation \Ref{HA} in Sec. V). 
Finally, the $U_{ab}(t)$ represent the freedom of the choice of the basis for the Hilbert space of the environment.

From the gauge freedom \Ref{eq-gauge-1}, 
we can always choose $H=\wt H$, $L_a=\wt L_a$, 
where a tilde denotes the traceless part of an operator, 
$\wt A:=A-(1/N)\Tr A$. 
Furthermore, by using the degrees of freedom of \Ref{eq-gauge-2}, 
we can also choose the $L_a$ to be mutually orthogonal. 
This is the gauge choice which we make throughout this paper.

\section{General Formalism}
\label{general formalism}
Let us consider the problem of controlling a certain physical system
governed by the master equation \Ref{master} and of steering its  
transition between given initial and final quantum states 
in the shortest time.
Mathematically this is a time optimality problem for the evolution of
the density operator $\rho(t)$ according to \Ref{master} by
controlling the Hamiltonian $H(t)$ and the Lindblad operators $L_a(t)$.

We assume that at least the `magnitudes' of the Hamiltonian and of the Lindblad
operators are bounded. 
Physically this corresponds to the fact that one can afford only a  
finite energy in the experiment, and that a finite level of decoherence is tolerated.
Besides this normalization constraint, 
the available operations may be subject also to other constraints, which
can represent either experimental requirements (e.g.,  the
specifications of the apparatus in use) or theoretical conditions (e.g.,
allowing no operations involving three or more qubits). 
We then consider the following action for the dynamical variables
$\rho(t)$, $H(t)$ and $L_a(t)$
\begin{align}
  \label{eq-action}
  S(\rho,H,L_a,\sigma,\lambda_a)=&\int dt\left[
  L_T+L_M+L_C\right],
\end{align}
with
\begin{align}
  \label{eq-L}
  &L_T:=
  \sqrt
  {\frac{g_{\rho}(\dot \rho,\dot \rho)}
    {g_{\rho}(\mathcal{L}(\rho),\mathcal{L}(\rho))}},
  \\
  &L_M:=\Tr[{\sigma \paren{\dot\rho-\LL(\rho)}}],
  \\
  &L_C:=
    \sum_{a}{\lambda_{a}}f_a(H) 
    +\sum_{a, b}\mu_{ab}[\Tr(\da L_a L_b)- N\gamma_a^2\delta_{ab}], 
\end{align}
where the traceless Hermitian operator $\sigma$, the real functions $\lambda_a$ and 
the complex functions $\mu_{ab}=\mu_{ba}^*$ are Lagrange multipliers, and $\Tr \rho(t)=1$.
As mentioned in the previous section, we assume 
(even before taking variations of the action) that $H$ and $L_a$ are traceless \cite{trace}, while
the second term in $L_C$ is a constraint which 
ensures the normalization and the mutual orthogonality of the $L_a$ operators.

The $L_T$ term in the action \Ref{eq-action} gives the time duration 
for the evolution of $\rho(t)$.
The Riemannian metric $g_\rho$ above is assumed to belong to the family of monotone metrics on
the space of density operators given by 
\begin{eqnarray}
  \label{eq-g}
  &&g_\rho(A,B)  :=  \Tr[{A\,c(L,R) (B)}], 
\end{eqnarray}
where $L$ and $R$ are multiplication of $\rho$ from the
left and right, respectively, 
$c(x,y) :=1/\paren{y \eta\paren{{x}/{y}}}$, and 
$\eta$ is an operator-monotone function satisfying 
$t\eta(t^{-1})=\eta(t)$ 
\cite{morozowa,petz}. 
The equations of motion derived below actually do not depend on the particular choice of
$g_\rho$. In fact, they are the same for all $L_T$ which
are constant {\it on shell}\/, i.e., 
when the master equation holds (and one may even choose $L_T\equiv1$). 

The $L_C$ term in \Ref{eq-action} generates the constraints 
\begin{align}
  \label{eq-f}
  f_a(H)=0
\end{align}
and 
\begin{equation}
\label{noise}
\Tr (\da L_a L_b)=N\gamma_a^2\delta_{ab}
\end{equation}
for the Hamiltonian and the 
Lindblad operators. 
The $L_M$ term in \Ref{eq-action} guarantees
that ${\rho(t)}$ satisfies
the master equation \Ref{master}. 

The other equations of motion are derived in the following way.
From the variation of $S$ by $\rho$, with the help of \Ref{master}, 
we get the  adjoint master equation 
\begin{align}
  \label{adj_master}
  \dot{\sigma}'=-\wt{\da{\mathcal{L}}(\sigma')}
\end{align}
for $\sigma':=\sigma
+{\wt{c(L,R)(\mathcal{L}(\rho))}}/{g_{\rho}(\mathcal{L}(\rho), \mathcal{L}(\rho))}$, 
where the superoperator $\da\LL$ is defined by $\Tr [\da A\LL(B)]=\Tr [\da\LL(A)B]$ and its 
explicit form is 
$\da\LL(A)= i [H,A]+\sum_a(\da L_aA L_a-\frac12\{\da L_a L_a,A\})$. 
The variation of $S$ by $H$ and \Ref{master} imply
\begin{align}
  \label{eq-dH}
  F=-i [\rho,\sigma'],
\end{align}
where we have defined the operator 
\begin{align}
  \label{eq-F-def}
  F(H):=\pder {L_C}H, 
\end{align}
which will be important in the sequel. 
From the variation of $S$ by $\da L_a$, together with \Ref{master}, 
we obtain
\begin{align}
  \label{eq-dL-pre}
\text{traceless part of }\left [ \sigma' L_a\rho-\frac{1}{2}L_a\{\rho,\sigma'\}\right ]=\sum_{b}\mu_{ab}L_b.
\end{align}
One can show that $\mu_{ab}$ is actually diagonal, i.e. that $\mu_{ab}=\mu_a\delta_{ab}$ ~\cite{diagonal}, 
and therefore get the algebraic formula
\begin{equation}
\text{traceless part of }\left [\sigma' L_a\rho-\frac{1}{2}L_a\{\rho,\sigma'\}\right ]=\mu_{a} L_a,
\label{eigen}
\end{equation}
which is an eigenvalue equation with eigenvalues $\mu_a$ for the eigenvectors $L_a$.
Finally, combining \Ref{master}, \Ref{adj_master}, \Ref{eq-dH} and \Ref{eigen}, 
we can eliminate the Lagrange multipliers to obtain 
the fundamental {\it quantum brachistochrone equation} 
\begin{align}
\label{eq-fund}
  i\dot F=[H,F].
\end{align}

We can thus obtain the time optimal $H(t)$ and $L_a(t)$ separately. 
One first solves \Ref{eq-fund} to obtain the optimal $H(t)$ and
then solves  \Ref{master}, \Ref{adj_master}, \Ref{eq-dH} and \Ref{eigen} to find the optimal $L_a(t)$. 
The quantum brachistochrone equation \Ref{eq-fund} is the same universal equation as for pure
states ~\cite{paper1} and for unitary operations ~\cite{paper2}.  
It can also be obtained by observing that the
first two terms $L_T$ and $L_M$ in the action \Ref{eq-action} are invariant under an arbitrary
infinitesimal non-Abelian transformation 
\begin{align}
  \label{eq-non Abelian gauge transf}
  &\rho\mapsto \rho - i [\Chi(t),\rho],
  \nn
  &\sigma\mapsto \sigma+i[\Chi(t),\sigma],
  \nn
  &H\mapsto H+\dot{\Chi}(t) - i [\Chi(t),H],  
   \nn
  &L_a\mapsto L_a- i [\Chi(t),L_a], 
\end{align}
where $\Chi(t)\in \su(N)$.
This is because \Ref{eq-non Abelian gauge transf} does not change the master equation \Ref{master} and the adjoint  master equation \Ref{adj_master}.
Therefore the variation of the entire action
reduces to the variation of the constraint term $L_C$, which is easily checked to produce the quantum brachistochrone equation \Ref{eq-fund}.
The same derivation also holds for the case of pure states and unitary operations treated in our previous works 
\cite{paper1,paper2}.

\section{A one-qubit example}
\label{onequbit}

Let us now discuss as an explicit example a one-qubit model where the Hamiltonian is subject only to the normalization constraint
\begin{equation}
\label{normalization}
f_0(H):=\frac{1}{2}(\Tr H^2 -N\omega^2)=0,
\end{equation}
where $\omega$ is a given constant.
In this case $F=\lambda_0 H$ and the quantum brachistochrone equation \Ref{eq-fund} becomes trivial, giving $H={\rm const}$ and $\lambda_0={\rm const}$ (see Sec. III in \cite{paper2}). 
Our problem then reduces to that of solving the master equation \Ref{master} and the adjoint master equation  \Ref{adj_master}
together with the algebaric equations \Ref{eq-dH} and \Ref{eigen} for $\rho(t),  L_a(t)$ and $\sigma'(t)$.
In the Pauli basis $\{\sigma_x,\sigma_y,\sigma_z\}$ these can be rewritten as equations for three-dimensional vectors.
Namely, if we parametrize  the states as
\begin{eqnarray}
  \rho(t)&:=&\frac{1}{2}[1+\vv{r}(t)\cdot\vv{\sigma}]\label{systemstate},
  \\
\sigma'(t)&:=&\vv{s}(t)\cdot\vv{\sigma},
\end{eqnarray}
and the Hamiltonian and the Lindblad operators as
\begin{eqnarray}
  H&:=&\vv{h}  \cdot\vv{\sigma},
  \\
L_a(t)&:=&\vv{l}_a(t)\cdot\vv{\sigma},
\label{H-L}
\end{eqnarray}
where $\vv{r}, \vv{s}$ and 
$\vv{h} \in \mathbb{R}^3$ and $\vv{l}_a \in \mathbb{C}^3$, 
the master equation \Ref{master} and the adjoint master equation \Ref{adj_master} can be rewritten as
\begin{eqnarray}
\!\!\!\! \dot{\vv{r}}&\!\!=&\!\!2[\vv{h}\times\vv{r}+\!\!\sum_{\vv{l}\in\{\vv{l}_a\}}
  (\Re((\vv{l}\cdot\vv{r})\vv{l}^*)\!-|\vv{l}|^2\vv{r}\!+i\vv{l}\times\vv{l}^*)],\label{master_v}\\
\!\! \!\!\dot{\vv{s}}&\!\!=&\!\!2[\vv{h}\times\vv{s}-\!\!\sum_{\vv{l}\in\{\vv{l}_a\}}
  (\Re((\vv{l}\cdot\vv{s})\vv{l}^*)-|\vv{l}|^2\vv{s})].\label{adj_master_v}
\end{eqnarray}
Moreover, the algebraic equations \Ref{eq-dH} and \Ref{eigen} read
\begin{equation}
 \vv{r}\times\vv{s}=\lambda_0\vv{h}={\text const}
\label{hamiltonian_v}
\end{equation}
and 
\begin{equation}
 K(\vv{r},\vv{s}) \vv{l}_a \!\!=\nu_a \vv{l}_a,
 \label{eigen_v}
\end{equation}
where $K(\vv{r},\vv{s})$ is the self-adjoint matrix
\begin{equation}\label{matrixK}
 K_{jk}(\vv{r},\vv{s}):=r_{j}s_{k}+r_{k}s_{j}-2i\sum_{l=x, y, z}\epsilon_{jkl}s_l
\end{equation}
and $\nu_a:=2(\vv{r}\cdot \vv{s}+\mu_{a})$ is a real number.

Because of the constraints \Ref{normalization} and \Ref{hamiltonian_v},
the components of the Hamiltonian $\vv{h}$ are given by 
\begin{equation}\label{ham}
 \vv{h}=
\biggl\{
\begin{array}{cc}
\pm\omega \frac{\vv{r}\times\vv{s}}{|\vv{r}\times\vv{s}|}
&\text{if}~~\vv{r}\times\vv{s}\ne \vv{0},
\\
\omega\vv{n}&\text{if}~~\vv{r}\times\vv{s}= \vv{0},
\end{array}
\end{equation}
where $\vv{n}$ is an arbitrary unit vector.
The components of the Lindblad operators $\vv{l}_a$ are determined as
eigenvectors of the eigenvalue equation 
(\ref{eigen_v})
with the constraints \Ref{noise}, i.e. $|\vv{l}_a|=\gamma_a$.

At a given instant, we parametrize $\vv{r}(t)$ and $\vv{s}(t)$ as
\begin{eqnarray}
 \vv{r}(t)&:=&r\left (\cos\frac{\theta}{2}\vv{e}_3+\sin\frac{\theta}{2}\vv{e}_1\right ),\\
 \vv{s}(t)&:=&s\left (\cos\frac{\theta}{2}\vv{e}_3-\sin\frac{\theta}{2}\vv{e}_1\right ),
\end{eqnarray}
where $\{\vv{e}_1(t), \vv{e}_2(t), \vv{e}_3(t)\}$ are orthonormal
 vectors, so that
 $\vv{r}\cdot\vv{s}=rs\cos\theta$,
with $r\in [0,1]$ and 
$\theta \in [0,\pi]$.
We can then rewrite the matrix $K$ in \Ref{matrixK} as
\begin{equation}\label{eigen_v_matrix}
(K_{jk})=2s\left[\begin{array}{ccc}
-r\sin^2\frac{\theta}{2} & -i\cos\frac{\theta}{2}&0\\
 i\cos\frac{\theta}{2}   &0  &i\sin\frac{\theta}{2}\\
0&-i\sin\frac{\theta}{2}&r\cos^2\frac{\theta}{2}
\end{array}
\right].
\end{equation}

\begin{figure}[h]
 \begin{center}
 \resizebox{6cm}{!}{\includegraphics{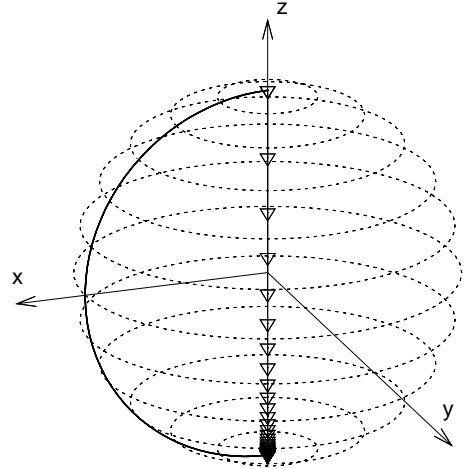}}
  \caption{Analytical, time optimal evolution of $\rho(t)$ (arrows) in the Bloch sphere 
 for the case of  $\vv{r}\times\vv{s}=\vv{0}$,  $\gamma_+\ne 0$, $\gamma_-=0$ and $\rho(0)=\ket{\uparrow}\bra{\uparrow}$.
  Also shown (thick solid meridian curve) the optimal pure state evolution between the north and south poles \cite{paper1}. }
  \label{NtoS}
 \end{center}
\end{figure}

If the conserved vector satisfies
$\vv{r}\times\vv{s}=\vv{0}$,
we can see $\theta=0$ (i.e. $\vv{r}=r\vv{e}_3$),
and the components of the 
Lindblad operators are given by (\ref{eigen_v_matrix}) as the following vectors
\begin{eqnarray}
 \vv{l}_{\pm}(t)&=&\frac{\gamma_{\pm}}{\sqrt{2}}(\vv{e}_1\pm i\vv{e}_2), \label{clindblad1}\\
 \vv{l}_0(t)&=&\gamma_0 \vv{e}_3. \label{clindblad2}
 \end{eqnarray}
To simplify the analysis,  we can move to the interaction
picture by the transformation $\rho'=U_0 \rho U_0^\dagger :=\frac{1}{2}(1\!+\!\vv{r}'\cdot\vv{\sigma})$ with
$U_0(t)=\mathcal{T}\exp(-i\int H(t)\dd t)$, so that the master equation
\Ref{master_v} 
becomes
\begin{equation}
 \dot{\vv{r}}'=-2[(\gamma_+^2+\gamma_-^2)\vv{r}'-(\gamma_+^2-\gamma_-^2)\vv{e}_3],
\end{equation}
which guarantees $\dot{\vv{e}}_3=\vv{0}$.
Therefore we obtain the following solution for the Bloch vector in the
interaction picture
\begin{equation}
\label{evol}
 \vv{r}'(t)=\left [\frac{\gamma_+^2-\gamma_-^2}{\gamma_+^2+\gamma_-^2}+
 \left (r'(0)-\frac{\gamma_+^2-\gamma_-^2}{\gamma_+^2+\gamma_-^2} \right )
 e^{-2(\gamma_+^2+\gamma_-^2)t}\right ]\vv{e}_3.
\end{equation}
If the magnitudes of the Lindblad operators are equal
i.e. $\gamma_+=\gamma_-$, the state will irretrievably lose the coherence,
but the coherence can be recovered when the magnitudes of the Lindblad
operators are different.
Note that in this particular case the $L_0$ operator, which corresponds to a projective measurement along $\vv{e}_3$, is not effective for the state evolution while the amplitude damping $L_{\pm}$ play
a significant role in \Ref{evol}.
The case $\vv{h}=\vv{0}$ is depicted in Fig. \ref{NtoS}.
In the case when $\vv{r}\times \vv{s}\not = 0$ the coupled equations \Ref{master_v}, \Ref{adj_master_v} and \Ref{eigen_v} can be solved numerically. We depict a family of optimal trajectories from a mixed state to a pure state in the Bloch sphere in Fig. 2.
 \begin{figure}[h]
 \begin{center}
\resizebox{6cm}{!}{\includegraphics{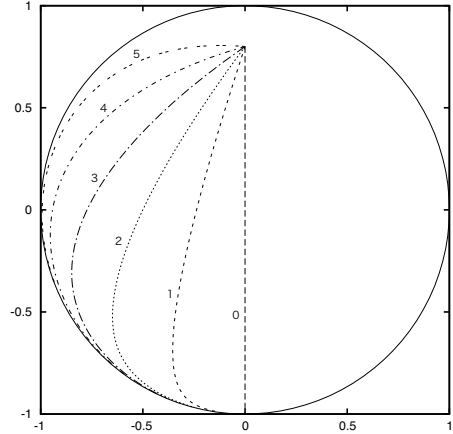}}
  \caption{Time optimal evolutions of a mixed state governed by the master
  equation with the Lindblad operators having magnitudes
  $(\gamma_1,\gamma_2,\gamma_3)=(1,0,0)$ in the descending order of the
  eigenvalues of \Ref{eigen_v}.
  Curves starting from $\vv{r}=(0,0,0.8)$ 
  and approaching $\vv{r}=(0,0,-1)$ are
  trajectories of the Bloch vector
  in the {\it x-z} plane cross-section of the Bloch sphere.
The initial direction of the curves is different for each initial angle
  between $\vv{r}$ and $\vv{s}$,
  i.e. $\vv{r}\cdot\vv{s}=rs\cos\frac{n\pi}{6}, n=0,1,..,5$.
}
  \label{numerical-mixed}
 \end{center}
\end{figure}

As a final remark for this section, we would like to point out that, in order to make the time duration of the transition physically well-defined, one can introduce a small but finite error region around the target state.
That is, one can be more interested in reaching the target state with a fixed  
fidelity close to one.
Then, for example, while mathematically it takes an infinite time for
the system to 
reach the target state, the system actually approaches the target state and enters into
its small surrounding region in a finite time.
We will show in Sec. VI  that in some cases the time optimal mixed state evolution can be faster than the time optimal pure state evolution (thick solid meridian curve in Fig. 1) discussed in \cite{paper1}.

\section{A model for measurement or decoherence}
\label{twoqubits}

 To get a further physical insight into our formalism, we study a simple
 model consisting of two interacting spins, one of which is identified
 with our system ($A$) and the other is an externally  controllable
ancilla ($B$). The extra ancilla spin can be regarded as representing either a measurement apparatus or an environment.  The two-qubit Hamiltonian can be chosen without loss of generality
(modulo local rotations of the system and ancilla qubits) as
\begin{equation}
H_{AB}(t):=\sum_{j, k=x,y,z} h_{jk}(t)\sigma_j\otimes\sigma_k,
\label{HAB}
\end{equation}
with time-dependent, tunable couplings $h_{jk}(t)$ \cite{niskanen}.
We simulate the optimal Hamiltonian and Lindblad operators discussed in the
previous section by tuning the couplings in \Ref{HAB} and the ancilla state. 
The state of the system
is described by the density operator \Ref{systemstate} 
and the state of the ancilla by
\begin{equation}
\rho_B(t):=\frac{1}{2}[1+\vv{b}(t)\cdot\vv{\sigma}],
\label{bloch}
\end{equation}
where $\vv{b}$ is a tunable Bloch vector.
The master equation in the Lindblad form \Ref{master}
comes from the repetition of the unitary evolution by the interaction $H_{AB}$ for a short  time duration $\tau$ (i.e., for $\tau$ much smaller than the typical dynamical timescale of $\rho(t)$) and the partial trace over
the $B$ state. Namely,
\begin{eqnarray}
\rho(t+\tau)&=&\Tr _B[e^{-i H_{AB}\tau }\rho(t) \otimes \rho_B(t)e^{iH_{AB}\tau}]\nn
&=&\rho(t)-i\tau \Tr _B[H_{AB},\rho(t) \otimes \rho_B(t)]\nn
&+&\tau^2 \biggl \{ \Tr_B[H_{AB}(\rho(t) \otimes \rho_B(t))H_{AB}]\nn
&&~~~-\Tr_B[\frac{1}{2}\{H_{AB}^2,\rho(t) \otimes \rho_B(t)\}] \biggr \}\nn
&+&\mathcal{O}(\tau^3).
\end{eqnarray}
For our model Hamiltonian we can perform the partial trace over the $B$ state and get
\begin{eqnarray}
\rho(t+\tau)&=&\rho(t)-i\tau [H,\rho(t)]\nn
&+&\tau^2\sum_{j, k=x,y,z} a_{jk}(t)\biggl [\sigma_j\rho(t)\sigma_k
-\frac{1}{2}\{\sigma_k\sigma_j,\rho(t) \}\biggr ]\nn
&+&\mathcal{O}(\tau^3),
\label{2model}
\end{eqnarray}
where the effective single qubit Hamiltonian $H$ is given in terms of the couplings $ h_{jk}$ and the Bloch vector $b_{j}$ as
\begin{equation}
H(t):=\sum_{j, k=x,y,z} h_{jk}b_{k}\sigma_j,
\label{HA}
\end{equation}
while the Lindblad matrix is defined by
\begin{equation}\label{aij}
a_{jk}(t):=\sum_{l, m=x, y, z} h_{jl}h_{km}\left(\delta_{lm}-i\sum_{n=x, y, z} \epsilon_{lmn}b_n\right).
\end{equation}
The eigenvectors and the eigenvalues of \Ref{aij} together
define the components of the Lindblad operators.
Therefore, since the same components are also the
eigenvectors in \Ref{eigen_v}, one has to
impose the commutativity condition $[a, K]=0$,
where $K$ is given by \Ref{matrixK}.
The real and imaginary parts of the latter condition give three and nine
constraints, respectively, so that we have twelve constraints in total
for the nine parameters of $h_{jk}$ and the three parameters of $b_{j}$.
Then we
can adjust the Hamiltonian couplings  $h_{jk}$ and the components $b_j$ of the
Bloch vector  of the ancilla state so that the optimal 
$\vv{r}(t), \vv{s}(t), \vv{h}(t)$ and $\vv{l}_a(t)$ obtained in the previous
section are reproduced.

In particular, if  $\vv{r}$ and $\vv{s}$ are parallel (and chosen to be along the {\it z}-axis as
at the end of the previous section), we can see that it is sufficient to
choose the coupling matrix $h_{jk}$ as
\begin{equation}\label{h_matrix}
(h_{jk})=
 \left[\begin{array}{ccc}
0 & p & 0\\
p & 0 & 0\\
0 & 0 & q
\end{array}
\right],
\end{equation}
with real numbers $p,q$, and the Bloch vector $\vv{b}$ as
\begin{equation}\label{b_vector}
\vv{b}=b\vv{e}_3.
\end{equation}
The Lindblad matrix \Ref{aij} becomes
\begin{equation}\label{a_matrix}
(a_{jk})=
 \left[\begin{array}{ccc}
  p^2 & ibp^2  &0\\
-ibp^2& p^2    &0\\
0     &  0     & q^2
\end{array}
\right].
\end{equation}
Then the $K$ matrix \Ref{eigen_v_matrix} for $\theta=0$,
\begin{equation}\label{K_matrix}
(K_{jk})=
2s \left[\begin{array}{ccc}
0 & -i & 0\\
i &  0 & 0\\
0 &  0 & r
\end{array}
\right],
\end{equation}
and the Lindblad matrix $a$ can be diagonalized simultaneously. 
In particular, the eigenvalues of $a$ are $p^2(1\mp b),q^2$ and the
corresponding eigenvectors are given as in \Ref{clindblad1} and
\Ref{clindblad2},
with the magnitudes of the Lindblad operators
\begin{eqnarray}
 \gamma_{\pm}^2&=&(1\mp b)\tau p^2,
 \\
 \gamma_0^2&=&\tau q^2.
\end{eqnarray}

The optimal solution for $\vv{r}(t)$ can be finally cast in the form
\begin{equation}
\label{damping}
 \vv{r}(t)=[-b+(r(0)+b)e^{-4p^2\tau t}]\vv{e}_3.
\end{equation} 
An intuitive explanation for the behavior of the Bloch vector is the following.
The exponential damping in \Ref{damping} can be attributed to the
standard formula for
the transition probability calculated from the interaction Hamiltonian
\begin{equation}\label{HAB2}
H_{AB}=p(\sigma_x\otimes\sigma_y+\sigma_y\otimes\sigma_x)+q\sigma_z\otimes\sigma_z.
\end{equation}
Note that $q$ does not appear in \Ref{damping} so that it can be put
equal to zero.
Suppose now that the $B$ spin is up, i.e. $b=1$. Since the interaction
Hamiltonian \Ref{HAB2} is proportional to
$\sigma_{+}\otimes\sigma_{+}-\sigma_{-}\otimes\sigma_{-}$, both the
system and the ancilla spins tend to align down.
Suppose we start with the completely mixed state $r(0)=0$.
In the statistical interpretation of the density operator,
this means that in half of the systems the spin is up while in the other
half the spin is down.
The up
components are steered down while the down components remain unchanged, so that
eventually the state of the system evolves from a completely mixed state
to a purely down state.
This kind of `cooling' has been already
discussed before \cite{sklarztannorkhaneja}.
On the other hand, for $b=0$, the system approaches a
completely mixed state regardless of the initial conditions, as we
expect because the system is kept into contact with the completely
random state of the ancilla, $\rho_B=1/2$.
This corresponds to decoherence.
We would like to remark that, although the coupling parameters $h_{jk}$ and
the ancilla state are constant, in our particular 
parallel $\vv{r}$ and $\vv{s}$ case, they can be time dependent in general. 
This particular demonstration of a simple two
qubit model suffices to illustrate the physics behind the master
equation in the Lindblad form.

\section{Optimal unitary evolution plus final measurement}
\label{closedsystem}

So far we have considered the situation in which the evolution of mixed states is the result of a series of short time unitary operations each followed by a measurement.
However, the evolution of mixed states can be also described by a finite time unitary evolution in an enlarged Hilbert space for the original system plus an environment followed by the trace over the environment, which is equivalent to a final measurement only. 
In this section we are going to study in more details this single measurement strategy and the dependence of the optimal duration time of the evolution on the size of the environment.

To be more explicit, let us consider a set of $n$-qubits in which the first qubit is
regarded as the system and the rest is the environment.
The whole closed system is then governed by the Schr\"odinger
equation with a controllable Hamiltonian $H(t)$ 
and the evolution of the first qubit can be expressed as
\begin{equation}\label{trace-out}
 \rho(t)=\Tr_B \{ U(t)\rho(0)\otimes \rho_B U^\dagger(t)\},
\end{equation}
where $\rho_B$ is the initial state of the environment and
$U(t)=\mathcal{T}\exp(-i\int H(t)\dd t)$.

Let us assume that there are no other restrictions for the system except the normalization
conditions \Ref{noise} and \Ref{normalization} for the Lindblad and the Hamiltonian operators, respectively.
Now, to justify the dependence of \Ref{normalization} for the traceless Hamiltonian on 
the dimension $N$ (for $n$ qubits, $N=2^n$) of the Hilbert space over which it acts, we note that
\begin{equation}
\label{normalization1}
\frac{1}{N}\Tr H^2 =\omega^2
\end{equation}
is required by the invariance under the trivial extension 
$H\mapsto H\otimes I_{M}$,
i.e.  $\frac{1}{MN}\Tr(H\otimes I_M)^2=\frac{1}{N}\Tr H^2$.
This is because the physical properties of the system described by $H$
should not change even if we add an extra ancillary environment 
of arbitrary dimension $M$ which is not interacting with the system.
A simliar argument applies to the Lindblad operators.
In particular, the constraints should be invariant under the trivial
extension of the Hilbert space. 
This simple observation is crucial to obtain the correct relation between the size 
of the environment and the optimal duration time of the evolution.
The same point will also become important in addressing the efficiency of quantum computation
and its scaling properties when we apply our formalism to explicit quantum 
algorithms \cite{paper4}.

Therefore,
the quantum brachistochrone equation \Ref{eq-fund} leads to a time independent
optimal Hamiltonian for the whole closed system which, without any loss of generality, can be chosen as
\begin{equation}
 H=\sqrt{2^{n-1}}\omega ( \ket{0}\bra{2^{n-1}}+\ket{2^{n-1}}\bra{0}),
\end{equation}
where the states $\ket{0}$ and $\ket{2^{n-1}}$ are the states $\ket{\uparrow}\otimes\ket{\uparrow}\otimes...\otimes\ket{\uparrow}$ and
$\ket{\downarrow}\otimes\ket{\uparrow}\otimes...\otimes\ket{\uparrow}$,
 respectively, in the binary representation.
The time evolution of the whole system is driven by the unitary matrix
\begin{eqnarray}
\label{U}
 U(t)&=&\cos(\sqrt{2^{n-1}}\omega t)
(\ket{0}\bra{0}+\ket{2^{n-1}}\bra{2^{n-1}})\nn
&-&i\sin(\sqrt{2^{n-1}}\omega t)
(\ket{0}\bra{2^{n-1}}+\ket{2^{n-1}}\bra{0}).
\end{eqnarray}
Suppose now that we want to steer the first qubit from the state $\rho(0)=\ket{\uparrow}\bra{\uparrow}$ up to the target state $\ket{\downarrow}\bra{\downarrow}$.
Then, assuming that $\rho_B=(\ket{\uparrow}\bra{\uparrow})^{\otimes n-1}$, from \Ref{trace-out} and \Ref{U} we obtain
\begin{eqnarray}
 \rho(t)=\frac{1}{2}\biggl [
1&+&\cos (2\sqrt{2^{n-1}}\omega t)(\ket{\uparrow}\bra{\uparrow}-\ket{\downarrow}\bra{\downarrow})\nn
 &+&\!\! i\sin(2\sqrt{2^{n-1}}\omega t)
(\ket{\uparrow}\bra{\downarrow}-\ket{\downarrow}\bra{\uparrow})\biggr ],
\label{wave}
\end{eqnarray}
so that the time required to reach the target state is 
\begin{equation}
\label{time}
T=\frac{\pi}{2\sqrt{2^{n-1}}\omega},
\end{equation}
which is faster than the pure state case,
where the time-optimal evolution from $\ket{\uparrow}$ to $\ket{\downarrow}$ under the same
constraint takes the time $T=\frac{\pi}{2\omega}$ \cite{paper1}.
We conclude the section by stressing the fact that the optimal duration time \Ref{normalization1} is an exponentially decreasing function of the number of the ancilla qubits of the environment.

\section{Summary and Discussion}
\label{summary}

We have developed a framework based on the variational principle for finding the time optimal quantum
operation to make a transition between given initial and final states $\rho_i$ and $\rho_f$, 
when the physical system obeys a Markovian master equation in the Lindblad form. 
The equations of motion for the Hamiltonian $H$ and the Lindblad operators $L_a$ 
can be written down once the constraints for $H$ and $L_a$ are specified according to the problem. 
One then obtains the time optimal operation $(H(t),L_a(t))$ and the optimal duration time $T$ 
by solving the quantum brachistochrone and the other equations of motion and imposing the initial and final conditions $\rho(0)=\rho_i$ and $\rho(T)=\rho_f$. 

Our formalism for the mixed state case has been explicitly demonstrated with a one-qubit model. 
First, the optimal Hamiltonian was obtained from the quantum brachistochrone equation.
Then the time evolution of the density operator and of the Lindblad operators which represent an optimal measurement or decoherence was found from a remaining set of ordinary differential and algebraic equations.
In a particular case an analytical solution was given, while the solution for more general situations was shown numerically. 
To get a more physical intuition, we constructed an interacting two-qubit model where an
ancilla qubit plays the role of the environment and we demonstrated that repeated
short-time Markovian transitions can reproduce the optimal time evolution of mixed states.  
Next we considered the time optimal evolution of mixed states driven by a single final measurement 
after the unitary evolution in an enlarged Hilbert space.
In this case the optimal duration time is an exponentially decreasing function of the number of the qubits of the environment.

Let us compare the efficiency of the time optimal evolutions between given initial and final pure states for the two models of repeated Markovian measurements and that of a final measurement following a unitary evolution in an extended Hilbert space.
\begin{figure}[h]
 \begin{center}
 \vskip 0.75cm
 \resizebox{7cm}{!}{\includegraphics{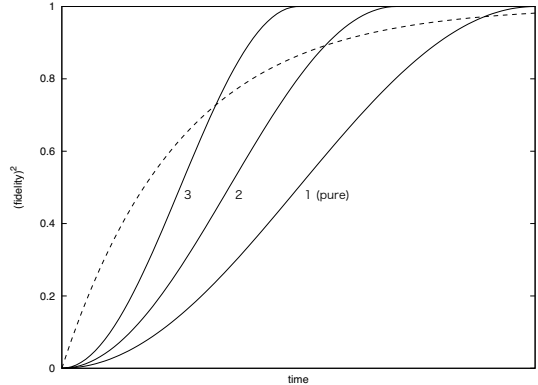}}
  \caption{Time dependence of the fidelity squared $\bra{\downarrow}\rho(t)\ket{\downarrow}$.
  The dashed and the solid curves correspond to 
the repeated Markovian measurement case \Ref{damping} and
the single, final measurement case \Ref{wave} in an $n$-qubit system ($n=1,2,3$), respectively.
Until the two curves cross each other, the repeated Markovian measurement
model approaches the target $\rho_f=\ket{\downarrow}\bra{\downarrow}$ faster than the single
measurement model.}
  \label{cos_exp}
 \end{center}
\end{figure}
In Fig. 3 we plot the fidelity between the target pure state and the time optimal evolved mixed state as a function of the duration time of the evolution.
We can see that, for a given fidelity close to one, the evolution via a final measurement becomes more and more efficient (i.e., it takes a shorter time) than the evolution via a repetition of measurements as the number of qubits in the environment increases.
This is because with an environment more resources are available for the processing of the information
required to generate the time optimal evolution.  

Incidentally, we note that both the dynamical evolution law \Ref{master} (or, equivalently, \Ref{2model}) and \Ref{trace-out} can be also expressed as a completely positive, trace preserving map $V(t)$, i.e. $\rho(t)=V(t)\rho(0):=\sum_{a=0}^{N^2}W_{a}(t)\rho(0)W_{a}^{\dagger}(t)$, where the $W_a$ are the Kraus operators, which satisfy $\sum_{a=0}^{N^2}W_{a}^{\dagger}W_{a}=1$.
In particular, the relation between the Lindblad operators in \Ref{master} and the Kraus operators is explicitly given by $W_{0}=1-\frac{\tau}{2}\sum_{a=0}^{N^2}L^{\dagger}_aL_a$ and $W_a=\sqrt{\tau}L_a$
(see, e.g., \cite{petruccione}).

Our work has not dealt with the more general case of 
different duration times for the contacts between the system and the environment and the case of the possible memory feed-back from the environment itself.
The authors of \cite{sklarztannorkhaneja} also considered the problem of control in dissipative quantum dynamics in order to achieve optimal purification of a quantum state, but they worked within the standard framework of a set of constant Lindblad operators.
Furthermore, although there should be no conceptual difficulty in extending our work to the problem of 
optimal quantum control via quantum feedback by introducing a stochastic term 
in the master equation \cite{belavkin,wisemandoherty,manciniwiseman}, we
have not discussed this problem here.

\section*{ACKNOWLEDGEMENTS}
This research was partially supported by the MEXT of Japan, under grant No. 09640341 (A.H. and T.K.), by the JSPS with grant L05710 (A.C.), by the COE21 project on `Nanometer-scale Quantum Physics' at Tokyo Institute of Technology (Y.O.).

\bibliographystyle{alpha}

\begin{thebibliography}{1}
\bibitem{trace} Our results are unchanged if one explicitly introduces Lagrange multiplier terms in the action \Ref{eq-action} in order to ensure the trace conditions for the operators $H$, $L_a$, $\sigma$ and $\rho$.
\bibitem{diagonal} With the help of the orthogonality condition \Ref{noise} we can see from \Ref{eq-dL-pre} that 
$N\mu_{ab}^2(\gamma^2_a-\gamma^2_b)=\Tr[\left (\sigma' L_b\rho-\frac{1}{2}L_b\{\rho,\sigma'\}\right)\da L_a - 
L_b\left (\sigma' L_a\rho-\frac{1}{2}L_a\{\rho,\sigma'\}\right)]$.
The right hand side of the latter formula is identically zero so that one can conclude that $\mu_{ab}=0$ if $\gamma_a=\gamma_b$. For equal $\gamma_a$, one can still rotate (using \Ref{eq-gauge-2}) the $L_a$ and diagonalize $\mu_{ab}$.
In all cases, we obtain $\mu_{ab}=\mu_a\delta_{ab}$.
\end{thebibliography}

\end{document}